\documentclass[aps,prc,preprint,groupedaddress,showpacs,showkeys]{revtex4}
\usepackage{amsmath,amssymb} 
\usepackage{dsfont}

\usepackage[dvips,dvipsnames]{color}
\usepackage[dvips]{graphicx}
\DeclareGraphicsExtensions{ps,eps}

\newcommand{\Balpha}{\ensuremath{\boldsymbol{\alpha}}}

\newcommand{\Br}{\ensuremath{\mathbf{r}}}
\newcommand{\Bx}{\ensuremath{\mathbf{x}}}
\newcommand{\Bq}{\ensuremath{\mathbf{q}}}

\DeclareMathOperator{\Tr}{Tr}

\begin{document}

\title{The $\alpha$ particle as a canonically quantized multiskyrmion}

\author{A.~Acus and E.~Norvai\v sas}
\email[]{acus@itpa.lt,     norvaisas@itpa.lt}
\affiliation{VU Institute of Theoretical Physics and Astronomy,
Go\v stauto 12, Vilnius 01108}
\affiliation{Department of Physics and Technology, Vilnius Pedagogical 
University, Lithuania}

\author{D.O.~Riska}
\email[]{riska@pcu.helsinki.fi}
\affiliation{Helsinki Institute of Physics, POB 64,
00014 University of Helsinki, Finland}

\thispagestyle{empty}
\date{\today}
\begin{abstract}

The rational map approximation to the solution to the
SU(2) Skyrme model with
baryon number  $B=4$ is canonically quantized. 
The quantization
procedure leads to anomalous breaking of the chiral symmetry, and
exponential falloff of the energy density of the soliton at large 
distances. The model is extended to SU(2) representations of 
arbitrary dimension. 
These soliton solutions capture the double node feature of the 
empirical $\alpha$ particle charge form 
factor, but as expected lead to a too
compact matter distribution. Comparison to phenomenology
indicates a preference for the fundamental representation.

\end{abstract}

\maketitle

\section{Introduction}

The chiral topological soliton model developed by Skyrme \cite{Skyrme61}, 
which represents a dynamical realization of the large $N$ limit of
QCD, describes many of the key qualitative features of 
of baryons and nuclei \cite{nyman,makhankov,manton}. The model 
describes baryons and nuclei as spatially extended topologically 
stable solitons of the
chiral meson field. The soliton solutions of the equation of motion 
are characterized by the winding number or topological charge of the
mapping $S^3\rightarrow S^3$, which is interpreted as the baryon number $B$. 
Numerical study has
shown that the shape of the ground state field configuration for 
nuclei with $B>1$ has an intriguing 
geometrical structure \cite{battye}. For $B=2$ the ground
state solution is toroidal and for $B=4$ the structure it
is octahedral.
Higher baryon number solutions are associated with more complicated 
symmetric polyhedral shapes. Such shapes also appear as
variational solutions to
interaction part of the nuclear Hamiltonian \cite{vijay}.

The rational map (RM) ansatz proposed for the SU(2) Skyrme model in 
\cite{Houghton98} provides a remarkably accurate analytic approximation
to the ground state solution of
the model. This ansatz  preserves the essential symmetries 
of the numerical solutions of the exact Skyrme model equations. The 
identification of the topological number with baryon number 
also leads to solitonic fullerene structures in light atomic nuclei 
\cite{Battye2001}. The RM ansatz has been generalized 
to the SU(3) Skyrme model as well \cite{Kopeliovich2000}.

The rational map ansatz for the SU(2) skyrmion for $B=2$,
which represents the deuteron, has been canonically
quantized in Ref.~\cite{Acus2004}
for representations of arbitrary dimension of the Skyrme model 
Lagrangian. The canonically quantized deuteron solutions and their
physical characteristics depend on the dimension of the
representation in contrast to the semiclassically quantized
solution. 

The matter density of the canonically quantized skyrmion soliton falls 
off exponentially
at long range in contrast to the power law falloff
of the classical soliton without a pion mass
term \cite{Acus2004,Acus98}. In the case of the $B=1$ skyrmion the 
inverse of the length scale of this
exponential falloff for corresponds to the pion mass, which
arises because of the anomalous breaking of chiral symmetry
by the canonical quantization procedure \cite{Acus98}. In the case
of the $\alpha-$particle it should correspond 
to $2\sqrt{m E_0}$, where $M$ is 
the nucleon mass and $E_0$ is the binding energy \cite{gibson}. Numerical 
calculation shows that the RM approximation leads to
exponential falloff at a somewhat smaller rate than this.
This feature may be traced to the fact that the Skyrme
model represents a large $N$ approximation to QCD, in which
the kinetic energy term for the nucleons vanish. The
ground state solution to the Skyrme model therefore
corresponds to the variational solution to the interaction
part of the nuclear Hamiltonian, as the kinetic energy
terms vanish in the large $N$ limit. 

Below the static observables and the
charge form factor of $^4$He are calculated from the
the quantum solution of the $B=4$ skyrmion obtained with the rational
map in 
SU(2) representations of arbitrary dimension. The calculated
charge form factor has the same two-node
structure as the experimental form factor, but the two
zeros appear at smaller values of momentum transfer than
in the empirical form factor. This shows that the ground
state solution of the Skyrme model has an unrealistically
compact structure, as expected. It
is instructive to compare the results to those previously
obtained with the product ansatz for the soliton field
\cite{boffi}. The product ansatz describes the asymptotic
long range 4-skyrmion structure of the solution, and leads
to a charge form factor for $^4$He, where nodes of the
calculated form factor in contrast occur at too large values of 
momentum
transfer. It is then natural to conjecture, that as the empirical
form factor is bracketed by the form factor calculated with the
too compact rational map approximation and with the too 
extended asymptotic
product ansatz, a more realistic solution of the Skyrme model
might provide an adequate description of the observed form
factor.

The organization of this paper is the following.
In section~\ref{sec2} the RM ansatz for the classical soliton 
of octaedral symmetry 
is generalized
to representations of arbitrary dimension. 
In section~\ref{sec3} canonical quantization of the
soliton is developed in the  
collective coordinate approach.
The numerical results for the properties of the
quantized solution are compared to the observables
of ${}^4$He in section~\ref{sec4}. Finally a
summarizing discussion is given in section~\ref{sec5}.

\section{The classical soliton of octaedral symmetry}
\label{sec2}
The Skyrme model is a Lagrangian density for a unitary
field $U({\bf x,t})$ that belongs to the representation of $SU(2)$ group.
In a general reducible representation  
$U({\bf x},t)$ may be expressed as a direct sum of Wigner's 
$D$ matrices for irreducible representations as:
\begin{equation}
U({\Bx},t)=\sum_j \oplus D^j(\Balpha ({\Bx},t))\,.
\label{a1}
\end{equation}
The $D^j$ matrices are functions of three unconstrained Euler angles 
$\Balpha=(\alpha ^1,\alpha ^2,\alpha ^3)$.

The chirally symmetric Lagrangian density of the Skyrme
model has the form:
\begin{equation}
{\cal L}\bigl(U({\Bx},t)\bigr)=
-{\frac{f_\pi ^2}4}{\Tr}\{R_\mu R^\mu \}+
{\frac 1{32e^2}}{\Tr}\{[R_\mu ,R_\nu ]^2\}\, .
\label{LagrangianDensity1}
\end{equation}
Here the ''right'' current  is defined as:
\begin{equation}
R_\mu =(\partial _\mu U)U^{\dagger },
\end{equation}
and $f_\pi $ (the pion decay constant) and $e$ are parameters. 

The rational map ansatz \cite{Houghton98} is an approximation
to the ground state solution of the Skyrme model with baryon 
number $B>1$ takes the following
form in a representation of arbitrary dimension:
\begin{equation}
U_{R}(\mathbf{r})=\exp (2\mathrm{i}\, \hat{n}^{a}\hat{J}_{(a)}F(r))\, .
\label{GeneralizedAnzats}
\end{equation}
Here $\hat{J}_{(a)}$ are SU(2) generators in a given 
representation. The unit vector $\hat{\mathbf{n}}$ may be defined
in terms of a rational complex function $R(z)$ as:
\begin{equation}
\hat {\mathbf n}_R=\frac{1}{1+|R|^2} \{2 \Re(R), 2 \Im(R), 1-|R|^2\}\, .
\end{equation}
For baryon number $B=4$ the function:
\begin{equation}
R(z)= \frac{z^4+2\sqrt{3} \mathrm{i} z^2 +1}{z^4-2\sqrt{3} \mathrm{i} z^2 +1}
\, ,
\label{RationalMapDef}
\end{equation}
has been found to be a suitable choice \cite{Houghton98}. Here $z=
\tan (\theta/2) \mathrm{e}^{\mathrm{i}\varphi}$ is a
complex coordinate that is parametrized by azimuthal and polar 
angles $\theta$ and $\varphi$. The circular
components of the unit vector $\hat {\mathbf n}_R$ are:  
\begin{eqnarray}
\hat{n}_{+1} &=&-\frac{1}{\sqrt{2}}  +
 \frac{\sqrt{3} \sin^2 \theta \bigl(\sqrt{3} \sin^2\theta  
-\mathrm{i}( 1 + \cos^2 \theta )\cos 2\varphi \bigr)}
  {2 \sqrt{2} \bigl( 1 - \sin^2 \theta + \sin^4 
\theta(1 - \sin^2 \varphi + \sin^4 \varphi )  \bigr) }
,\notag \\[2pt]
\hat{n}_{0} &=&\frac{\sqrt{3} \sin^2 \theta \cos \theta \sin 2\varphi}
  {1 - \sin^2 \theta + \sin^4 \theta(1 - \sin^2 \varphi + \sin^4 \varphi )},
\\[2pt]
\hat{n}_{-1} &=&\frac{1}{\sqrt{2}}  +
\frac{\sqrt{3} \sin^2 \theta \bigl(-\sqrt{3} \sin^2\theta 
 -\mathrm{i}( 1 + \cos^2 \theta )\cos 2\varphi \bigr)}
  {2 \sqrt{2} \bigl( 1 - \sin^2 \theta + \sin^4 \theta(1 - 
\sin^2 \varphi + \sin^4 \varphi )  \bigr)}.\notag
\end{eqnarray}

The rational map (\ref{RationalMapDef}) has cubic symmetry. The
orientation is fixed below so that the z-direction is that
of the third component of the angular momentum.

Differentiation of $\hat{\mathbf{n}}$ yields the relation  
\begin{equation}
(-1)^s(\nabla_{-s} r\, \hat n_m )(\nabla_{s} r\,\hat n_{m^\prime})=
\hat n_m \hat n_{m^\prime} +\mathcal{I} \bigl((-1)^m 
\delta_{-m,m^\prime}-\hat n_m \hat n_{m^\prime} \bigr),
\end{equation}
which proves to be useful in the explicit calculation of Lagrangian density 
\eqref{LagrangianDensity1}.
Here $\nabla_{s}$ are the circular components of the
nabla operator.  The symbol $\mathcal{I}$ here denotes the
function: 
\begin{equation}
\mathcal{I}=\Bigl(\frac{1+|z|^2}{1+|R|^2} \left|
\frac{\mathrm{d} R}{\mathrm{d} z}\right|\Bigr)^2\, ,
\end{equation} 
the explicit form of which is:
\begin{equation}
\mathcal{I}=\frac{12 \sin^2 \theta \,(1 - \sin^2 \theta + 
\sin^4 \theta\, \sin^2\varphi\, \cos^2 \varphi)}{\bigl( 1 - 
       \sin^2 \theta + \sin^4 \theta\, ( 1 - \sin^2 \varphi +
 \sin^4 \varphi ) \bigr)^2}\, .
\end{equation}
Integrals of powers of $\mathcal{I}$ over $\theta$ and $\phi$ can be regarded 
as Morse functions \cite{Houghton98}.

The baryonic charge density takes the following
form in the irrep $j$ :
\begin{equation}
\mathcal{B}(r,\theta,\varphi)=\varepsilon^{0 k \ell m} \Tr R_k R_\ell R_m \\
=-8\, j(j+1)(2 j+1) \mathcal{I}\,\frac{F^\prime(r) \sin^2 F}{r^2}\, .
\label{BarionicDensity}
\end{equation}
Because of the presence of the $\mathcal{I}$ function in this 
expression, there is no need to modify usual boundary conditions 
$F(0)=\pi;\quad F(\infty)=0$ for the chiral angle. The baryon 
number therefore takes the standard expression:
\begin{equation}
B=\frac{1}{24 N \pi^2}\int_0^\infty\mathrm{d}r 
\int_0^{2\pi}\mathrm{d}\varphi\int_0^\pi \mathrm{d}\theta\
 \mathcal{B}\ r^2 \sin\theta \, , 
\label{BarionicChargeDef}
\end{equation}
with the normalization factor $N=\frac23 j(j+1)(2j+1)$, as 
expected \cite{Acus98}. The normalization factor is chosen 
to be unity in the fundamental representation of SU(2). The present
choice of boundary conditions ensures that the integral of 
the $\mathcal{I}$ function is proportional to the baryon number:
\begin{equation}
\int_0^{2\pi}\mathrm{d}\varphi\int_0^\pi \mathrm{d}\theta\ 
\mathcal{I}\sin \theta  = 4\pi B \, .
\label{Idef}
\end{equation}

Substitution of the rational map  ansatz \eqref{GeneralizedAnzats} into 
the Lagrangian density \eqref{LagrangianDensity1} leads to the
classical Skyrme model density:
\begin{equation}
\begin{split}
&\mathcal{L}_{\textrm{cl}}(r,\theta,\varphi)=
-N \biggl(f_\pi^2\Bigl( \frac{F^{\prime 2}(r)}{2} +
\frac{\mathcal{I} \sin^2 F}{r^2}\Bigr)
+\frac{1}{e^2}\frac{\mathcal{I} \sin^2 F}{r^2} 
\Bigl( F^{\prime 2}(r) +\frac{\mathcal{I} \sin^2 F}{2 r^2}\Bigr)\biggr)\, .
\label{ClassicalLagrangianDensity}
\end{split}
\end{equation}
Note, that the symmetry of the Lagrangian density 
\eqref{ClassicalLagrangianDensity} in the $\theta, \varphi$ space is 
completely determined by the function $\mathcal{I}$ and its 
(more symmetric) powers.

It is useful to introduce dimensionless 
coordinates $\tilde r = e f_\pi r$.
Variation of Lagrangian then yields the following
differential equation for chiral angle:
\begin{eqnarray}
&&F^{\prime\prime}(\tilde r)\Bigl(1+\frac{2B\sin^2 F(\tilde r)}
{\tilde r^2}\Bigr)+\frac{2 F^\prime(\tilde r)}{\tilde r}+
\frac{F^{\prime 2}(\tilde r)B\sin 2F(\tilde r)}{\tilde r^2}
\nonumber\\
&&-
\frac{B \sin 2 F(\tilde r)}{\tilde r^2} -
\frac{I_2 \sin^2 F(\tilde r)\sin 2 F(\tilde r)}{\tilde r^4}=0.
\label{ClassicalModelEquation}
\end{eqnarray}
Here we have used the abbreviation:
\begin{equation}
I_2=\frac{1}{4 \pi}\int_0^{2\pi}\mathrm{d}
\varphi\int_0^\pi\mathrm{d}\theta \mathcal{I}^2 \sin\theta\, .
\end{equation}
In the limit $\tilde r \rightarrow \infty$, the equation 
\eqref{ClassicalModelEquation} reduces to simple asymptotic form
\begin{equation}
F^{\prime\prime}(\tilde r)+\frac{2 F^\prime(\tilde r)}{\tilde r}-
\frac{2 B F(\tilde r)}{\tilde r^2}=0\, .
\label{ClassicalModelAssympthoticsEquation}
\end{equation}
From this the asymptotic large distance solution, which satisfies
physical boundary conditions, can easily be obtained as:
\begin{equation}
F(\tilde r)= C_1 \tilde r^{-\frac{1+\sqrt{1+8 B}}{2}}\, .
\label{ClassicalAssympthotics}
\end{equation}
Here $C_1$ a constant to be determined later by continuous
joining of the numerical small distance solution onto the
analytic asymptotic solution.
For  $B=4$, the power of $\tilde r$ in \eqref{ClassicalAssympthotics} 
is $\approx - 3.37$. Note that eqs. 
(\ref{ClassicalLagrangianDensity}--\ref{ClassicalAssympthotics}) are 
valid for all $B$, provided that the corresponding function 
$\mathcal{I}$ is used.  

\section{Canonical quantization in the collective coordinate approach}
\label{sec3}

The quantization of the Skyrme model in a general representation
\cite{Acus98} can be carried out by means of
collective rotational coordinates that separate the variables,
which depend on the time and spatial coordinates \cite{Nappi83}: 
\begin{equation}
U(\Br,\Bq (t))=A\left(\Bq(t)\right) U_R(\Br) A^\dagger \left(
\Bq(t)\right)\, .  \label{CollectiveCoordinatesDef}
\end{equation}
Here the three real Euler angles $\Bq(t)=(q^1(t),q^2(t),q^3(t))$ are 
quantum variables. These are sufficient for the $\alpha$ particle
ground state, for which $S=T=0$. 

The canonical quantization with constraints 
procedure employed here was originally 
suggested by Dirac~\cite{Dirac1964}, and further developed 
in refs.~\cite{Kimura1972, Gitman1990}.
In this formalism the Skyrme Lagrangian \eqref{LagrangianDensity1} is 
considered quantum mechanically {\it ab
initio} in contrast to the conventional semiclassical quantization 
of the Skyrmion as a rigid body.
In the SU(2) case canonical quantization implies that the
three independent 
generalized coordinates $\Bq(t)$ and 
the corresponding velocities $\dot{\Bq}(t)$
satisfy the following commutation relations \cite{Fujii87}: 
\begin{equation}
\lbrack \dot q^a,\,q^b]=-\mathrm{i}f^{ab}(\Bq)\, .
\label{CommutationRelation1}
\end{equation}
Here $f^{ab}(\Bq)$ are functions of generalized coordinates 
$\Bq$ only, the explicit forms of which are determined 
self-consistently upon imposition of the
quantization condition. The tensor $f^{ab}$ is 
symmetric with respect to interchange of the indices $a$ and $b$ 
by the relation $[q^a,\,q^b]=0$. 

The commutation 
relation between a generalized velocity component $\dot q^a$ and 
an arbitrary function $G(\Bq)$ is given by: 
\begin{equation}
\lbrack \dot q_a,\,G(\Bq)]=-i\sum_rf^{ar}(\Bq)
{\frac \partial {\partial q^r}}G(\Bq)\, . 
\label{CommutationRelation2}
\end{equation}
Here Weyl ordering of the operators has been employed:
\begin{equation}
\partial_0 G(\Bq) =\frac12 \{\dot q^\alpha,
 \frac{\partial}{\partial q^\alpha} G(\Bq)\}\, .
\label{WeylOrdering}
\end{equation}
The curly brackets denote an anticommutator. With this choice of 
operator ordering no further ordering ambiguity appears.
 
To derive the Lagrangian the expression 
\eqref{CollectiveCoordinatesDef} is 
substituted into the Lagrangian density \eqref{LagrangianDensity1}. 
Consider first the term 
that is quadratic in the generalized velocities. After integration 
over the spatial coordinates the 
Lagrangian takes the form:
\begin{equation}
L(\Bq,\dot \Bq, F) =\frac{1}{N}\int {\rm d}^{3}\Br{\cal 
L}(\Br,\Bq(t),F(r))= \frac12 \dot q^\alpha 
g_{\alpha \alpha ^\prime} \dot q^{\alpha^\prime} +{\cal O}(\dot{q}^0)\, .
\label{LagrangianWith2Velocities}
\end{equation} 
Here the momentum of inertia tensor is:
\begin{equation}
g_{\alpha \alpha ^\prime}= C^{\prime (b)}_\alpha (\Bq) 
E_{(b) (b^\prime)} C^{\prime (b^\prime)}_{\alpha^\prime} (\Bq)\, .
\end{equation}
Here $E_{(b) (b^\prime)}$ is defined as:
\begin{equation}
E_{(b) (b^\prime)} =-\frac12 (-1)^b a_b(F) 
\delta_{b,-b^\prime}\quad \textrm{\small (no summation over $b$)}\, .
\label{TensorOfMomentaInertia}
\end{equation}
Here  $a_1=a_{-1}$.
The soliton momenta of inertia are given as 
\begin{equation}
\begin{split}
a_0(F)=\frac{\tilde a_0}{e^3 f_\pi} = 4\pi \int_0^\infty 
r^2\sin^2 F\Bigl((1-N_2) \bigl( f_\pi^2+\frac{1}
{e^2}F^{\prime 2}\bigr)+\frac23\frac{B}{e^2}
 \frac{\sin^2 F}{r^2} \Bigr)\,\mathrm{d}r\, , \\
a_1(F)=\frac{\tilde a_1}{e^3 f_\pi} = 2\pi
 \int_0^\infty r^2\sin^2 F\Bigl((1+N_2)
 \bigl( f_\pi^2+\frac{1}{e^2}F^{\prime 2}\bigr)+
\frac43\frac{B}{e^2} \frac{\sin^2 F}{r^2} \Bigr)\,\mathrm{d}r \, .
\end{split}\label{MomentaInertiaDef}
\end{equation}
The symbol $N_k$ in this expression denotes the angular integrals: 
\begin{equation}
N_k=\frac{1}{4\pi}\int_0^\pi \mathrm{d}\theta 
\int_0^{2\pi} \mathrm{d}\varphi \sin\theta \hat n_0^k.
\end{equation}
For baryon number $B=1$ and $B=2$ the integrals may be
evaluated in closed form to yield
$N_2(\textrm{nucleon})=\frac13$;
$N_4(\textrm{nucleon})=\frac15$ and
 $N_2(\textrm{deuteron})=-1+\frac{\pi}{2}, 
N_4(\textrm{deuteron})=-1/3 +\frac{\pi}{4}$. For $B=4$ the
numerical values of the corresponding integrals
are $N_2\approx 0.218897$ and $N_4\approx 0.118382$. 
The other integrals, which explicitly enters calculation of 
the inertia tensor\eqref{TensorOfMomentaInertia}, may be
evaluated analytically by the following
expression: 
\begin{equation}
\begin{split}
&\int \Bigl(\frac{1+|z|^2}{1+|R|^2}\Bigl| 
\frac{\mathrm{d}R}{\mathrm{d}z}\Bigr|\Bigr)^2
 \Bigl(\frac{1-|R|^2}{1+|R|^2}\Bigr)^m
\frac{2 \mathrm{i} \mathrm{d}z\mathrm{d}\bar z}{(1+|z|^2)^2}=
\int_0^\pi \mathrm{d}\theta \int_0^{2\pi} \mathrm{d}\varphi
 \sin\theta\,\mathcal{I} \hat n_0^m \\
&= 2\pi B\frac{(-1)^m+1}{m+1},\qquad m\in Rationals;\quad m\ge 0. 
\end{split}
\label{MorseIntegratingFactor}
\end{equation}
The validity of expression has been verified numerically 
for a number of randomly chosen rational maps with different 
baryon numbers $B$  
to a very high degree of precision. There is good reason to
conjecture that the integrals are topologically conserved 
quantities valid for all rational maps.
Note that the relation \eqref{Idef} is a particular 
case ($m=0$) of eq. \eqref{MorseIntegratingFactor}. Here the 
function $\mathcal{I}$ plays an intriguing role as 
an "integrating" factor.

The coefficients $C_{\alpha }^{\prime (b)}$ and their inverses $C^{\prime 
\alpha }_{(b)}$ are functions of the dynamical variables, which appear
 in the differentiation of the Wigner D matrices:
\begin{equation}
\frac{\partial}{\partial \alpha ^k} D^j_{m n} (\Balpha )=-\frac1{\sqrt{2}}
C^{\prime (a)}_k(\Balpha)D^j_{m m^\prime} (\Balpha ) \bigl\langle
 jm^\prime\bigl| J_{(a)}\bigr| j n\bigr\rangle \, .
\end{equation}

The conventional quantum mechanical commutation relations 
$\bigl[ p_\alpha , q^\beta \bigr]=-\mathrm{i} \delta_{\alpha \beta}$ 
for the momenta
$p_\alpha = \frac{\partial L}{\partial \dot q^\alpha}=
\frac12 \bigl\{\dot q^\beta, g_{\alpha \beta}\bigr\}$ then leads
to the following expression for the tensor $f^{\alpha\beta}$
(\ref{CommutationRelation1}): 
\begin{equation} 
f^{\alpha \beta}(\Bq)=g^{-1}_{\alpha \beta}(\Bq)\, . 
\end{equation}
It is convenient 
to introduce the following angular momentum operators on the 
hypersphere $S^3$ (the manifold of the SU(2) group):
\begin{equation}
\hat J^\prime _{(a)} = -\frac{\mathrm{i}}{\sqrt{2}} \bigl
\{p_\alpha , C^{\prime \alpha}_{(a)} \bigr\}\, .
\label{JPrimeDef}
\end{equation}
It is readily verified that the operator $\hat{J}_{a}^{\prime }$ 
is a $D^{j}(\Bq)$ ''right rotation'' generator that has the well
defined actions:
\begin{equation}
\hat J^{\prime 2} \left|
\arraycolsep=0.3\arraycolsep
\renewcommand{\arraystretch}{0.6}
\begin{array}{c}
\ell  \\ 
{m_s,m_t}
\end{array}
\right\rangle=\ell (\ell+1) \left| 
\arraycolsep=0.3\arraycolsep
\renewcommand{\arraystretch}{0.6}
\begin{array}{c}
\ell  \\ 
{m_s,m_t}
\end{array}
\right\rangle
;\qquad 
\hat J^{\prime 2}_0 \left| 
\arraycolsep=0.3\arraycolsep
\renewcommand{\arraystretch}{0.6}
\begin{array}{c}
\ell  \\ 
{m_s,m_t}
\end{array}
\right\rangle=m_t^2 \left| 
\arraycolsep=0.3\arraycolsep
\renewcommand{\arraystretch}{0.6}
\begin{array}{c}
\ell  \\ 
{m_s,m_t}
\end{array}
\right\rangle;
\end{equation}
on the normalized state vectors with fixed spin and isospin $\ell$:
\begin{equation}
\left| 
\arraycolsep=0.3\arraycolsep
\renewcommand{\arraystretch}{0.6}
\begin{array}{c}
\ell  \\ 
{m_s,m_t}
\end{array}
\right\rangle =\frac{\sqrt{2\ell +1}}{4\pi }D_{m_s,m_t}^{\ell }(\Bq
)\left| 0\right\rangle .  \label{StateVector}
\end{equation}

The explicit form of the function  $f^{ab}(\Bq)$, in turn, leads to 
to an explicit expression of the Skyrme model Lagrangian density
\eqref{LagrangianDensity1} in the collective coordinate approach. 
Lengthy manipulation and use of computer algebra \cite{Wolfram2003}
yields the result:
\begin{equation}
\begin{split}
&{\cal L}_{\textrm{qt}}(r)=-N\Biggl(
f_\pi^2\biggl\{ \frac{F^{\prime 2}}{2} + \frac{\mathcal{I} \sin^2 F}{r^2} 
-\frac{\sin^2 F}{8}\biggl[ \Bigl(\frac{1}{a_0} + \frac{3}{a_1}\Bigr)
 \mathds{C} - \frac{2}{a_1} \Bigl(\frac{1}{a_0} + \frac{1}{a_1}\Bigr)\\
&\hphantom{{\cal L}_{\textrm{qt}}(r)=-N\Biggl(f_\pi^2\biggl(
 \frac{F^{\prime 2}(r)}{2} +  }
+ \frac{(2 j - 1) (2 j + 3)  \sin^2 F}{5}\Bigl(3\mathds{C}^2 -
 \frac{4}{a_1}\mathds{C} + \frac{4}{a_1^2} \Bigr)\biggr] \biggr\}\\
&
+\frac{1}{e^2}\biggl[\frac{\mathcal{I} \sin^2 F}{r^2} \Bigl(F^{\prime 2}
 + \frac{\mathcal{I} \sin^2 F}{2 r^2}\Bigr)
-\frac{\sin^2 F}{8}\biggl(\frac{\mathcal{I} \sin^2 F}{r^2}
 \Bigl[\Bigl(\frac{1}{a_0}+ \frac{1}{a_1}\Bigr)\mathds{C}-
\frac{2}{a_0 a_1}\Bigr]\\
&
+F^{\prime 2} \Bigl[\Bigl(\frac{1}{a_0} + \frac{3}{a_1}\Bigr)\mathds{C}
 - \frac{2}{a_1} \Bigl(\frac{1}{a_0} + \frac{1}{a_1}\Bigr)\Bigr]
+
\frac{(2 j - 1) (2 j + 3)}{5}\Bigl\{-\frac{\mathcal{I} \sin^2 F}{r^2}
 \Bigl[ 3 \mathds{C}^2 \\
&-2 \Bigl(\frac{2}{a_0} + \frac{5}{a_1}\Bigr)\mathds{C} +\frac{2}{a_1}
 \Bigl(\frac{4}{a_0}+\frac{3}{a_1} \Bigr)\Bigr] 
+F^{\prime 2}\Bigr(
3\mathds{C}^2 - \frac{4}{a_1}\mathds{C} + \frac{4}{a_1^2} -
 2\mathds{C}^2\sin^2 F \Bigr)\Bigr\} 
\biggr)\biggr]\Biggr)\, .
\label{QuantumLagrangianDensity}
\end{split}
\end{equation}
Here the following notation has been introduced:
\begin{equation}
\mathds{C}=\frac{1}{a_0}+ \frac{1}{a_1}-\Bigl(\frac{1}{a_0}- 
\frac{1}{a_1}\Bigr)\hat n_0^2.
\end{equation}
The expression \eqref{QuantumLagrangianDensity} does not contain the
operator component.
Integration of the latter (operator component) yields
matrix elements, which depend on spin and isospin $\ell$:
\begin{equation}
\begin{split}
&\left\langle 
\arraycolsep=0.3\arraycolsep
\renewcommand{\arraystretch}{0.6}
\begin{array}{c}
\ell  \\ 
{m_s,m_t}
\end{array}
\right|
\int_0^\infty\mathrm{d}r\int_0^\pi \sin\theta\mathrm{d}\theta 
\int_0^{2\pi} \mathrm{d}\varphi \Bigl(f_\pi^2 +\frac{1}
{e^2}\bigl[\frac{\mathcal{I}}{r^2}\sin^2 F+F^{\prime 2}(r)\bigr]\Bigr)
\biggl(\frac{1}{a^2_1} \hat{\mathbf{J}}^{\prime 2}(q) 
\\
&\hphantom{\left\langle 
\arraycolsep=0.3\arraycolsep
\renewcommand{\arraystretch}{0.6}
\begin{array}{c}
\ell  \\ 
{m_s,m_t}
\end{array}
\right|}
+ \bigl(\frac{1}{a^2_0}-\frac{1}{a^2_1}\bigr) \hat{J}^{\prime 2}_0(q)
 - \Bigl[\frac{1}{a_1}\bigl(\hat{\mathbf{J}}^{\prime}(q) 
\cdot\mathbf{\hat n}\bigr) +\bigl(\frac{1}{a_0}-\frac{1}{a_1}\bigr)
 \hat{J}^{\prime}_0(q)\hat n_0\Bigr]^2\biggr)
\left| 
\arraycolsep=0.3\arraycolsep
\renewcommand{\arraystretch}{0.6}
\begin{array}{c}
\ell  \\ 
{m_s,m_t}
\end{array}
\right\rangle
\\
&\hphantom{\left\langle \quad \right|}
=m_t^2\bigl(\frac{1}{a_0}-\frac{1}{a_1}\bigr) +\frac{\ell(\ell+1)}{a_1} .
\end{split}
\end{equation}
This expression vanishes in the case of $^4$He for
which $m_t=\ell=0$. 

Integration and subsequent variation of Lagrangian density 
\eqref{QuantumLagrangianDensity} then leads to the following
integro-differential 
equation for the quantum chiral angle in the dimensionless coordinate 
$\tilde r = e f_\pi r$:
\begin{equation}
\begin{split}
&F^{\prime \prime}(\tilde r)\Biggl(4\tilde r^2 + 8 B \sin^2 F(\tilde r)
 + e^4 \tilde r^2 \sin^2 F(\tilde r) \biggl[4 \tilde \mu^2\\
&\hphantom{F^{\prime \prime}(\tilde r)\Biggl(4\tilde r^2 + 8 B \sin^2 
F(\tilde r)}
+ 
\frac{(2 j - 1) (2 j + 3)}{5} \Bigl(\mathds{A}  + 2 \mathds{B} +
 \bigl( \mathds{A} + \mathds{B}\bigr) \cos 2 F(\tilde r) \Bigr)\biggr]\Biggr)
\\
+&F^{\prime 2}(\tilde r)\Biggl(4 B \sin 2 F(\tilde r) + e^4 \tilde
 r^2 \sin 2 F(\tilde r) \biggl[2 \tilde \mu^2\\
&\hphantom{F^{\prime 2}(\tilde r)\Biggl(4 B \sin 2 F(\tilde r) + e^4}
+
\frac{(2 j - 1) (2 j + 3)}{10} \Bigl(\mathds{B} + 2 \bigl( \mathds{A}
  + \mathds{B}\bigr) \cos 2 F(\tilde r) \Bigr)\biggr]\Biggr)
\\
+&\tilde r F^{\prime}(\tilde r)\Biggl(8 + e^4 \sin^2 F(\tilde r)
 \biggl[8 \tilde \mu^2 + 
\frac{2 (2 j - 1) (2 j + 3)}{5} \Bigl(\mathds{A}+2 \mathds{B} +
 \bigl( \mathds{A}  + \mathds{B}\bigr) \cos 2 F(\tilde r) \Bigr)\biggr]
\Biggr)
\\
-&\sin 2 F(\tilde r) \Biggl( 4 B +\frac{4 I_2 \sin^2 F(\tilde r)
}{\tilde r^2}+
e^4 \tilde r^2  \biggl(2 \tilde \mu^2 + \frac{(2 j - 1) (2 j + 3)}{5}
 \bigl(2 \mathds{A}+3 \mathds{B}\bigr)\sin^2 F(\tilde r)\biggr)
\\
&+2 e^4 B \sin^2 F(\tilde r)\biggl\{2 \tilde \mu^2_{0}+
\frac{1}{3 \tilde a_1^2}+\frac{2}{3\tilde a_0 \tilde a_1}+
\frac{(2 j - 1) (2 j + 3)}{5}\biggl(-\frac{8}{15 \tilde a_0^2}+
\frac{6}{15\tilde a_0 \tilde a_1}-\frac{13}{15 \tilde a_1^2}\\
&
+ \frac{4\pi (-1 + 3 N_2)}{9}\Bigl(\frac{1}{\tilde a_0} -
\frac{1}{\tilde a_1}\Bigr)
\biggl[ \Bigl(\frac{3}{\tilde a_0^2}+\frac{2}{\tilde a_0 \tilde a_1}+
\frac{1}{\tilde a_1^2}\Bigr)\int \tilde r^2 \sin^2 2 F(\tilde r)
\mathrm{d} \tilde r\\
&\hphantom{+ \frac{4\pi (-1 + 3 N_2)}{9}\Bigl(\frac{1}{\tilde a_0}
 -\frac{1}{\tilde a_1}\Bigr)\biggl(}
+8 \Bigl(\frac{1}{\tilde a_0^2}+\frac{1}{\tilde a_0 \tilde a_1}+
\frac{1}{\tilde a_1^2}\Bigr)\int \tilde r^2 \sin^4 F(\tilde r)
 F^{\prime 2}(\tilde r) \mathrm{d} \tilde r
\biggr]\biggr)
\biggr\}\Biggr)\, .
\label{QuantumEquation}
\end{split}
\end{equation}
Here
\begin{eqnarray}
&&\mathds{A}=-\frac{4}{\tilde a^2_1} + \frac{4}{\tilde a_1}
 (-1 + N_2)\Bigl(\frac{1}{\tilde a_0} -\frac{1}{\tilde a_1}\Bigr)\, , \\
&&\mathds{B}=(-1 + 2 N_2 - N_4) \Bigl(\frac{1}{\tilde a_0}
 -\frac{1}{\tilde a_1}\Bigr)^2\, .
\label{ABShortcuts}
\end{eqnarray}
Above $\tilde \mu^2$ denotes the following integral:
\begin{equation}
\begin{split}
&\frac{4 \tilde \mu^2}{e^4} =\frac{(-1+4 m_t)(-1+N_2)}{\tilde a_0^2}
 -\frac{\tilde a_0 (1+N_2)}{\tilde a_1^3}
+\frac{2 \bigl(1+ (1+N_2) (1+m_t -\ell (\ell +1))\bigr)}{\tilde a_1^2}
\\
&+\frac{8\pi B}{3 \tilde a_1} \Bigl(\frac{2(-1+N_2)}{\tilde a_0^2}-
\frac{1+N_2}{\tilde a_0\tilde a_1}-\frac{1+N_2}{\tilde a_1^2}\Bigr)
\int \sin^4 F(\tilde r)\mathrm{d} \tilde r
\\
&
+\frac{(2 j - 1) (2 j + 3)}{5}\Biggl(
\frac{3(-1+N_2)}{\tilde a_0^2}\Bigl(N_4-5-\frac{2\tilde a_1 (-1 + N_4)}
{\tilde a_0}\Bigr)
\\
&\phantom{+}
+\frac{2(1+2N_2-3 N_4)}{\tilde a_0 \tilde a_1}+
\frac{1+N_2
}{2\tilde a_1^2} \Bigl(3 N_4+9+\frac{\tilde a_0 (1 + 3N_4)}{\tilde a_1}
\Bigr)
\\
&\phantom{+}
+16\pi\biggl( \frac{-1+N_2}{\tilde a_0^2}\Bigl(
\frac{1 -2 N_2+N_4}{\tilde a_0}
-\frac{-1 + N_4}{\tilde a_1}\Bigr)\\
&+\frac{1+N_2}{2\tilde a_1^2} \Bigl(\frac{-1+ N_4}{\tilde a_0}-
\frac{1 +2 N_2 + N_4}
{\tilde a_1}\Bigr)\biggr)\int \tilde r^2 \sin^4 F(\tilde r) 
F^{\prime 2}(\tilde r)\mathrm{d} \tilde r
\\
&\phantom{+}
+\frac{8\pi B}{15}\biggl( \frac{-1+N_2}{\tilde a_0^2}
\Bigl(\frac{-1 +45 N_4}
{\tilde a_0}-\frac{-31 +45 N_4}{\tilde a_1}\Bigr)
\\
&\phantom{+}
+\frac{1+N_2}{2\tilde a_1^2}\Bigl(\frac{-31 +45 N_4}{\tilde a_0}-
\frac{29 +45 N_4}
{\tilde a_1}\Bigr)\biggr)\int \sin^4 F(\tilde r)\mathrm{d} \tilde r
\\
&\phantom{+}
+2\pi\biggl( \frac{-1+N_2}{\tilde a_0^2}\Bigl(
\frac{3(1 -2 N_2+N_4)}{\tilde a_0}+
\frac{1 +2 N_2-3 N_4}{\tilde a_1}\Bigr)
\\
&\phantom{-}
-\frac{1+N_2}{2\tilde a_1^2} \Bigl(\frac{1+2 N_2-3 N_4}{\tilde a_0}+
\frac{3 +2 N_2
 + 3N_4}{\tilde a_1}\Bigr)\biggr)
\int \tilde r^2 \sin^2 2 F(\tilde r)\mathrm{d}
 \tilde r\Biggr)\, .
\end{split}
\label{MuDef}
\end{equation}
The symbol $\tilde \mu^2_{0}$ represents
the special case of $\tilde \mu^2$ integral, when
$N_2=\frac{1}{3}$. 

At large distances the eq.~\eqref{QuantumEquation} reduces to the 
asymptotic form:
\begin{equation}
\tilde r^2 F^{\prime\prime}(\tilde r) +2 \tilde r F^{\prime}
(\tilde r)-\bigl(2 B+ \tilde \mu^2 \tilde r^2\bigr) F(\tilde r)=0\, .   
\end{equation} 
From this asymptotic equation it follows that the quantity 
$\tilde \mu$ describes the falloff rate of the 
chiral angle at large distances:
\begin{equation}
F(\tilde r)=C_1\mathrm{e}^{-\tilde \mu \tilde r}
\Bigl(\frac{\tilde \mu}{\tilde r}+\frac{B}{\tilde r^2}\Bigr)\, .
\label{AsymptoticSolution}
\end{equation}
The related quantity $\mu=e f_\pi \tilde \mu$ describes the asymptotic 
falloff $\exp (-2 \mu r)$ of the soliton mass density for the dimensional
coordinate $r$. Note the appearance of  $\tilde \mu^2$ 
in all the higher derivative terms in eq.~\eqref{QuantumEquation}, 
which is a nontrivial result.
The equations (\ref{MomentaInertiaDef},
\ref{MorseIntegratingFactor},\ref{QuantumLagrangianDensity},
\ref{QuantumEquation} and \ref{MuDef})
are conjectured to be valid for all rational maps $R(z)$.

Because of the isospin of $^4$He is zero, the charge distribution is 
proportional 
to the baryon density~\eqref{BarionicDensity}.
The charge form factor then is the usual Fourier transform:
\begin{equation}
F_c(q)=\frac12 \int \mathrm{d}^3 j_0(q\, r) \mathcal{B}(r,\theta,\varphi),
\label{ChargeFF}
\end{equation}
where $j_0$ denotes the spherical Bessel function of zero order.

\section{Numerical results}
\label{sec4}
The RM ansatz represents an approximation,
which gives energies that fall
above the
numerically computed ground state energy
by only a few percent~\cite{Battye2001}.  
Calculation of the static properties and the charge form factor of 
${}^4$He from the RM with $B=4$ should therefore be expected
to give a good approximation to those for the exact ground 
state solution. In the present
numerical calculation the parameters
$f_\pi$ end $e$ in the Skyrme model Lagrangian were
determined so as to reproduce the calculated static
observables of the nucleons in the different representations $j$ 
considered in Ref.~\cite{Acus98}. 

The quantum integro-differential equation \eqref{QuantumEquation} was
solved numerically by the shooting method. In the initialization of
the algorithm trial values for all the integrals 
($a_0, a_1, \mu^2,\ldots$) that appear in the
equation had to be specified. For these estimates were obtained
by employment of the  
semiclassical chiral angle of the $B=1$ skyrmion. 
Shooting from the point $\tilde r_\textrm{max}$,
 (where $F(\tilde r)$ assumed to be of the form
\eqref{AsymptoticSolution}) 
to the point $\tilde r_\textrm{min}$ 
(here $F(\tilde r)= F(\tilde r_\textrm{min})- (\tilde r_\textrm{min}-\tilde r)
 F^\prime(\tilde r_\textrm{min})$) and varying the only unknown
constant $C_1$ in \eqref{AsymptoticSolution} leads to  
a continuous function $C^2$ that satisfies the
required boundary conditions $F(0)=\pi$
and $F(\infty)=0$.

Typically $\tilde r_\mathrm{max}\approx 6$ and 
$\tilde r_\mathrm{min}\approx 10^{-2}$ 
(the equation has a singularity at the origin). The
chiral angle function found by this method is then
used to recalculate all required integrals and procedure is 
iterated until the integrals converge to a stable value. 
The convergence proved to be rapid, and faster than in the case
of the nucleon. Every iteration increases the absolute integral 
precision approximately by one decimal point. 
Thus typically 10--15 iterations are enough to achieve 
an accurate solution for the chiral angle. 

\begin{figure}
\begin{center}
\includegraphics*[scale=0.6]{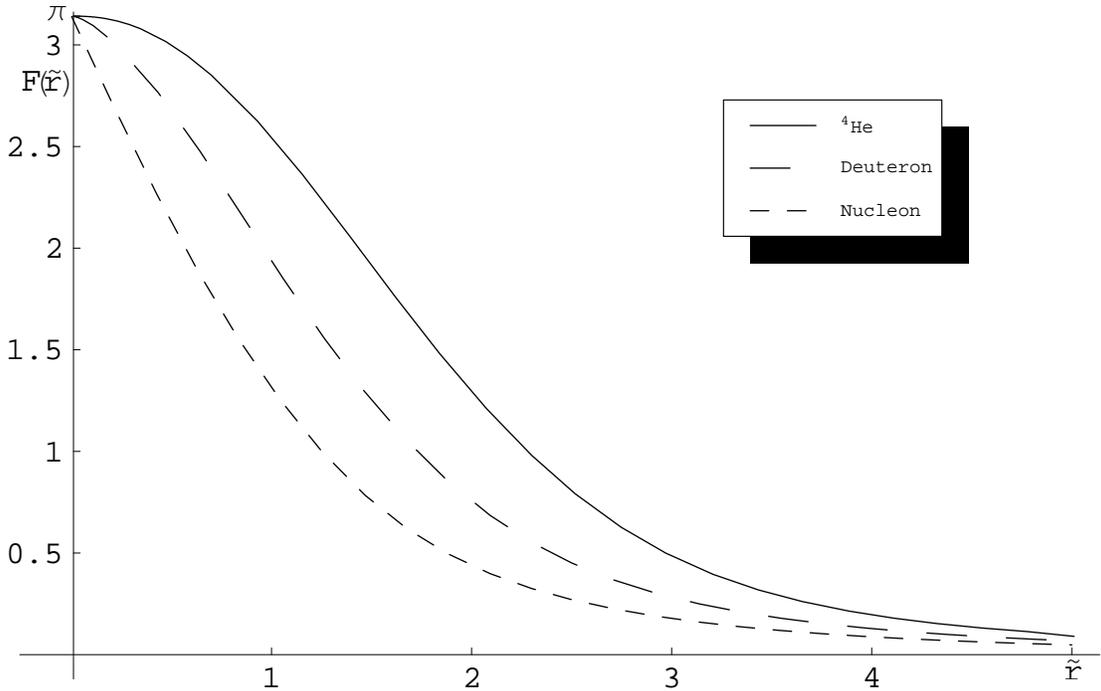}
\caption{${}^4$He (solid), deuteron (long-dashed) and nucleon
 (short-dashed) chiral angle profile functions in SU(2) representation
 $\frac12$.}
\label{chiralAnglesWithLegend}
\end{center}
\end{figure}

The canonical quantization of the $B=1$ skyrmion, which
describes the nucleon, was presented in Ref.~\cite{Acus98}.
Variation of the quantized energy functional 
revealed the existence of stable solutions
for the nucleon. In that work the
parameters were determined by the isoscalar radius ($0.72$~fm)
and mass ($939$~MeV) of the nucleon.
The same parameter values for $f_\pi$ and $e$ in the Skyrme
model Lagrangian were employed here for the solution
of the $B=4$ soliton, which describes the ${}^4$He nucleus
in different representations. 
Fig.~\ref{chiralAnglesWithLegend} shows 
the chiral angle profile functions
for different baryon numbers 1, 2 and 4~$B$. Here the
rational
map ans\"atze were used in the case of $B=2$ and $B=4$. It is 
notable that the exponential falloff
rate of the chiral angle becomes slower and smoother with
increasing baryon number.

The calculated values of
the static 
observables of ${}^4$He  
are listed in Table~\ref{table1}.
The best agreement between the calculated and the empirical
values for the
charge radius
$\langle r^2_E\rangle ^{1/2}$ and the corresponding
binding energy $E_0$ values is
 found for the reducible representation 
 $1\oplus\frac{1}{2}\oplus\frac{1}{2}$ as in the case of the 
nucleon~\cite{Acus98}. For the higher irreps 
no binding is found at all with these parameter values.

While the finite pion mass is conventionally
introduced by adding an explicitly chiral 
symmetry breaking pion mass term to the Lagrangian density of the 
model \cite{Nappi83}, the canonical quantization procedure by itself
gives rise to a finite
pion mass. This realizes Skyrme's original
conjecture that "This (chiral) symmetry is, however, destroyed by the 
boundary condition ($U(\infty )=1$), and we believe that the mass (of pion) 
may arise as a self consistent quantal effect" \cite{Skyrme62}. 

The "quantal effect" (the exponential
falloff rate of the mass density of ${}^4$He, 
$\mathrm{e}^{-2\mu r}$) which we find in  \eqref{AsymptoticSolution} is, however,
much smaller than the value that is
obtained for a 4-nucleon system with the empirical binding
energy: $\mu=\sqrt{ m E_0}$, where $m$ denotes
nucleon mass. The reason for this is that the rational map
ansatz gives an approximation to the ground state solution,
which does not contain the vibrational modes. 
This conclusion is also supported by comparison to the
semiclassical approximation to the $B=4$ skyrmion 
given in ref. \cite{walhout}, which did take into
account the vibrational modes, and obtained both
a smaller binding energy (79 MeV) and concomitantly a
larger radius (1.50 fm).
Alternatively
it may be viewed as natural consequence of the implied large
$N$ limit of the model, in which there is no kinetic energy
contribution from the constituent nucleons.

The nonrelativistic charge form factors which are calculated
from fixed empirical values of nucleon~\cite{Acus98} have the
same qualitative features as the empirical form factor values 
taken from Refs~\cite{Frosch67,Arnold78}, with two nodes. 
The best agreement with experimental data is
found for the fundamental representation $j=\frac12$.

\begin{table}
\caption{The predicted static ${}^4$He nuclei observables in different
representations with fixed empirical values for the nucleon isoscalar
radius $0.72$~fm and nucleon mass $m_N=939$~MeV~\cite{Acus98}.
The momenta of inertia, $\tilde a_i$, are in units of $1/(e^3 f_\pi)$.}
\begin{tabular}{c|c|c|c|c|c}
\hline
$j$&		$1/2$&	  $1$&	$3/2$&
	$\scriptscriptstyle 1\oplus\frac{1}{2}\oplus\frac{1}{2}$&
	\rm{Exp.} \\
\hline
\hline
$f_\pi$&	 59.8&  58.5 &  57.7 &   58.8 			
				&$93$ MeV \\
$e$  &		4.46 &  4.15 &  3.86 &   4.24 			
				&  \\ 
$m$&   3585   &  3759   &  3975   &  3701			
			&$3728.55$ MeV \\
$\mu $
	   &$33.1$    & $45.2$ &$50.4$  &$41.8$			
			&229? MeV \\
$\langle r^2_E\rangle ^{1/2}$
	   &$1.39$     & $1.52 $ &$1.65$   &$1.49$			
			&$1.676$ fm \\
$E_0$
	   &$-171$     & $+3$    &$+219$   &$-55$
						
&$-28.11$ MeV \\

$\tilde a_0$
	   &$157.1$     & $154.6$    &$152.9$   &$155.2$
						
& \\
$\tilde a_1$
	   &$130.1$     & $128.1$    &$126.8$   &$128.6$
						
&
\end{tabular}
\label{table1}
\end{table}

\begin{figure}
\begin{center}
\includegraphics*[scale=0.6]{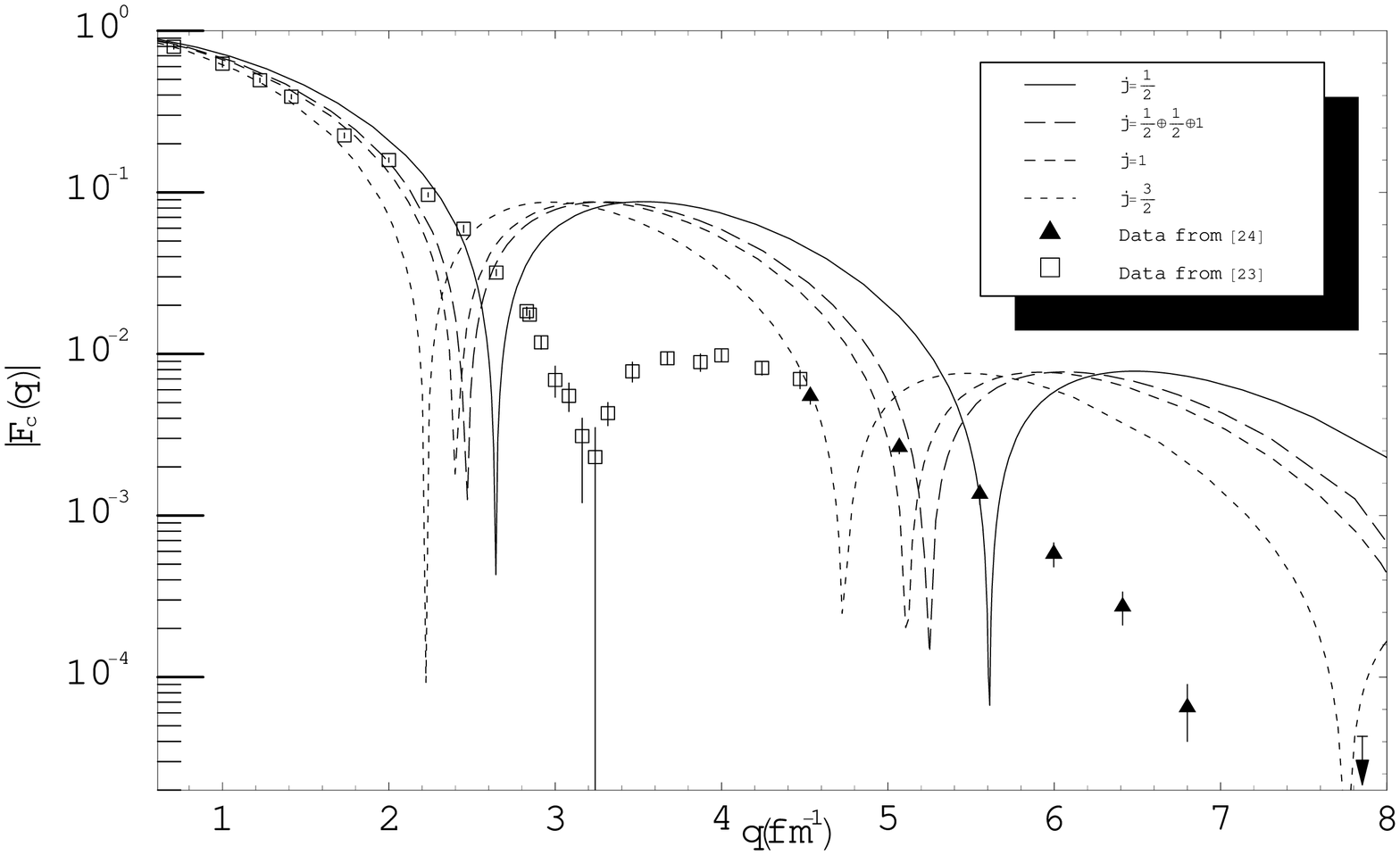}
\caption{Comparison of ${}^4$He electric form factors in different
 representations of SU(2) with experimental data~\cite{Frosch67,Arnold78}.
The form factors are calculated with parameters that yield
the experimental nucleon mass $m_N=939$~MeV
and radius $r=0.72$~fm~\cite{Acus98}}.
\label{formFactorsFromNucleonDataod}
\end{center}
\end{figure}

\section{Discussion}
\label{sec5}

The main result derived above is the demonstration of the
utility of the rational map approximation for the $B=4$
skyrmion, which allows the complete canonical quantization
of the soliton to be carried out in closed form in
a way similar to, even though calculationally more
cumbersome than, that used for the hedgehog solution
for the $B=1$ skyrmion. 

From the phenomenological
perspective the main result is however the explicit
demonstration that the empirical charge form factor
of the $\alpha-$particle is bracketed between the
form factor derived here by the rational map ansatz, which
approximates the ground state, and the form factor 
that is obtained with the product ansatz \cite{boffi},
and which asymptotically approaches the configuration
of 4 separated $B=1$ skyrmions. This then suggests that
there exists a smooth path between these two limiting
configurations, and that a physically more realistic
solution may eventually be found on this path. That
is yet another example of the remarkably wide field
of baryonic phenomenology, for which the Skyrme model
provides a qualitative description.

In the case of the $B=4$ skyrmion it was found that
the calculated observables in the fundamental representation 
lead to a better qualitative agreement with the 
empirical values than those obtained in representations
of larger dimension. In the case of the $B=1$ and $B=2$
skyrmions there is no such clear preference for the
fundamental representation \cite{Acus2004}.

The quantization of the deuteron (the $B=2$ skyrmion) is of particular
interest due to the different values of spin $S=1$ and isospin $T=0$.
This implies that quantization with three quantum variables as in
eq.~\eqref{CollectiveCoordinatesDef} is not sufficient. 
In ref.~\cite{Acus2004} six independent degrees of
freedom -- ie right and left chiral transformations were therefore 
employed. This allowed the construction of quantum states with different  
values of spin and isospin. Such a quantization of classical states 
with predefined symmetry, $\hat{\mathbf{n}}$ should be applicable 
to a wide class of 
nuclei.

As noted above the canonical quantization procedure generates a
pion mass term as originally conjectured by Skyrme \cite{Skyrme62}.
In work based on the conventional semiclassical quantization the 
pion mass term has in contrast to be introduced by way of an explicit 
chiral symmetry breaking
term. In that method the requirement of rotational stability 
requires a value for the pion mass that is considerably
larger than the empirical value \cite{Battye2005}. With
such large values for the pion mass the chiral symmetry breaking
term leads to spatial configurations for
the ground state solution of the Skyrme model with baryon number
larger than 4, which differ significantly from those obtained
in the chiral limit \cite{Battye2006}.  
It should be worthwhile to explore how the features implied by
the overly large pion mass term in the semiclassical approximation are
be modified once the mass term, which arises dynamically 
in consistent canonical quantization procedure are taken
into account. The canonical
quantization procedure here applied cannot, however, be directly applied
to this question as it keeps the angular dependence of the ansatz fixed.

\begin{acknowledgments}

E.~N. acknowledges the hospitality of the Helsinki
Institute of Physics during course of this work.
Research supported in part by the Academy of Finland grant
number 54038 and Lithuanian State Science and Studies Foundation grant 
number T-06034.

\end{acknowledgments}

\end{document}